\documentclass[12pt,preprint]{aastex}

\slugcomment{Accepted for publication in the Astrophysical Journal}

\shorttitle{Infrared Emission from HD\,189733b}
\shortauthors{Deming et al.}

\begin{document}



\title{Strong Infrared Emission from the Extrasolar Planet HD\,189733b}


\author{Drake Deming}
\affil{Planetary Systems Laboratory, NASA's Goddard Space Flight Center, 
Mail Code 693, Greenbelt, MD 20771}
\email{ddeming@pop600.gsfc.nasa.gov}

\author{Joseph Harrington}
\affil{Center for Radiophysics and Space Research, Cornell University, 
326 Space Sciences Bldg., Ithaca, NY  14853-6801}
\email{jh@oobleck.astro.cornell.edu}

\author{Sara Seager}
\affil{Department of Terrestrial Magnetism, Carnegie Institution of 
Washington, 5241 Broad Branch Rd NW, Washington, DC  20015}
\email{seager@dtm.ciw.edu}

\author{L.\ Jeremy Richardson\altaffilmark{1} }

\affil{Exoplanets and Stellar Astrophysics Laboratory, NASA's Goddard 
Space Flight Center, Mail Code 667, Greenbelt, MD 20771}
\email{richardsonlj@milkyway.gsfc.nasa.gov}


\altaffiltext{1}{NRC Postdoctoral Research Fellow}


\begin{abstract}
We report detection of strong infrared thermal emission from the
nearby ($d=19$ pc) transiting extrasolar planet HD\,189733b, by
measuring the flux decrement during its prominent secondary eclipse.  A
6-hour photometric sequence using Spitzer's infrared spectrograph in
peak-up imaging mode at $16~\mu$m shows the secondary eclipse depth
to be $0.551\pm0.030\%$, with accuracy limited by instrumental
baseline uncertainties, but with $32\sigma$ precision ($\sigma =
0.017\%$) on the detection. The $16~\mu$m brightness temperature of
this planet ($1117\pm42$K) is very similar to the Spitzer detections
of TrES-1 and HD\,209458b, but the observed planetary flux
($660~\mu$Jy) is an order of magnitude greater. This large signal
will allow a detailed characterization of this planet in the infrared.
Our photometry has sufficient signal-to-noise ($\sim400$ per point) to
motivate a search for structure in the ingress/egress portions of the
eclipse curve, caused by putative thermal structure on the disk of the
planet.  We show that by binning our 6-second sampling down to $\sim
6$-minute resolution, we detect the modulation in the intensity
derivative during ingress/egress due to the overall shape of the
planet, but our sensitivity is not yet sufficient to distinguish between
realistic models of the temperature distribution across the planet's
disk.  We point out the potential for extending Spitzer secondary
eclipse detections down to the regime of transiting hot Neptunes, if
such systems are discovered among nearby lower main sequence stars.
\end{abstract}




\section{Introduction}

The detection of infrared (IR) thermal emission from two extrasolar planets
\citep{charb05, deming} using the {\it Spitzer Space Telescope}
\citep{werner} opened a new era in which planets orbiting other stars
can be studied directly.  The Spitzer detections were made by
observing the flux decrement during secondary eclipse in transiting
systems.  The recently discovered planet transiting the star
HD\,189733 \citep{bouchy} is particularly suitable for Spitzer
detection and characterization, because it transits a relatively small
star -- allowing maximum planet-star contrast -- and because the
distance to the system is only 19 pc \citep{perryman}.  HD\,189733b has
an orbital period of only 2.219 days \citep{bouchy}, putting it in the
class of `very hot Jupiters'.  \citet{gaudi} suggested that the very
hot Jupiters are a separate dynamical class of exoplanets.  Since the
other members of this class orbit much fainter stars, the discovery of
HD\,189733 may allow a previously impossible direct comparison between
different classes of extrasolar planets.

In this paper we report detection of a prominent secondary eclipse of
HD\,189733b using Spitzer observations at $16~\mu$m.  We confirm that
the strong IR thermal emission from this planet will indeed permit
detailed characterization studies.  To begin such studies, we
examine our data for structure in the ingress/egress portions of the
eclipse curve, as can be caused by temperature structure on the disk
of the planet.  We thus attempt the first exploratory observations of
spatially-resolved structure on the disk of a planet orbiting another
star.

\section{Observations}
\label{sec:obs}

Whereas the first Spitzer detections \citep{charb05, deming} were made
using the IRAC and MIPS instruments, we here use the Infrared
Spectrograph (IRS, \citealp{houck}) in the peak-up imaging (PUI) mode
to detect HD\,189733b.  This mode provides imaging photometry at a
wavelength ($16~\mu$m peak, $\sim 5~\mu$m~FWHM) intermediate between
the $8~\mu$m IRAC and $24~\mu$m MIPS bands.  Our PUI photometry began
on 17 November 2005 at 23:54 UTC.  We placed the star alternately at
two positions on the detector array, separated by 25\arcsec (13.5
pixels).  We obtained 15 six-second exposures of the star at each
position, then nodded to the other position, and repeated this cycle
98 times.  We thus acquired a total of 1470 images during the 6-hour
observation.  The nod procedure allowed us to examine the zodiacal
background at each position, out of phase with the stellar
observations. This permits us to check the flat-fielding, using the
spatially uniform background. The nod also permits measurement of
latent image effects, and it provides insurance against unanticipated
hot or inoperative pixels.

\section{Analysis}
\label{sec:analysis}

\subsection{Photometry}

The $16~\mu$m zodiacal background in our observations of HD\,189733 is
approximately 9 MJy/sr, and this is sufficiently weak compared to
HD\,189733 (peak intensity $\sim 250$ MJy/sr) to allow simple aperture
photometry, not limited by background noise.  After eliminating 5
images having obvious flaws, we summed the intensity for each image in
a $9 \times 9$-pixel box centered on the star, and subtracted the
background level.  The background level for each image was determined
from a histogram of the pixel values outside of the stellar box,
fitting a Gaussian to determine the centroid of the
histogram. Photometry from the two nod positions differed by a
constant factor close to unity ($1.005$), but showed no other
differences above the noise.  We normalized the measurements at
both nod positions so as to yield the same average intensity.

We also performed aperture photometry on 2MASS20004297+2242342, which
is $11\arcsec$ distant, and about 4 magnitudes fainter than
HD\,189733.  We set the boundary of the $3 \times 3$-pixel box for
this comparison star at the intensity minimum between the overlapping
PSF wings of both stars. The comparison star contributes a very small
flux in the photometry aperture for HD\,189733 ($\sim 0.5\%$), and we
corrected for this using a modeled Spitzer PSF.

Figure~1 shows the stellar photometry vs. time for HD\,189733, before
background subtraction.  We record about $10^6$ electrons in each
exposure, of which $\sim 6.3 \times 10^5$ are from the star.  We
therefore expect the stellar SNR to be $\frac{6.3 \times 10^5}{10^3} =
630$.  The point-to-point scatter in our final photometry is $\sim
0.0025, SNR \sim 400$. We were not able to improve this significantly
by decorrelating against other parameters. For example, we found no
significant correlation between the photometric noise and fluctuations
in the position of the star on the detector (typically $\sim 0.05$
pixels, $\sim 0.1$ arc-sec).

The SNR of our photometry is $\sim~60$\% of the photon noise
limit, which is more than sufficient to detect the secondary eclipse
of the planet. The eclipse is already obvious as the dip in the time
series on Figure~1.  A vertical line on the figure at phase 0.5 marks
the nominal center of the eclipse.  The reality of the eclipse is
established by the fact that it is not present in either the
comparison star or the background time series.  Moreover, the
amplitude, central phase, and shape of the eclipse are in close accord
with expectations.

\subsection{Baseline Fitting}

Figure~1 shows that the measured intensity of the star increases
steadily over the 6-hour observation sequence.  Both the comparison
star and the background level show a similar increase,
which we denote as the `ramp'.  There are two peak-up arrays in IRS,
and the red ($22~\mu$m) array (that was pointed to adjacent sky) shows
a similar ramp.  This ramp is a previously unreported instrument effect,
not yet understood by the IRS instrument team.  In July 2005 we
observed two secondary eclipses of HD\,209458 using IRS in the
$7-14~\mu$m spectroscopic mode.  During these spectroscopic
observations, the peak-up arrays (always operating) were both
observing adjacent sky, and the backgound in all cases exhibits a
similar ramp.  We verified that the ramp appears in raw data, so it
cannot be an artifact of the pipeline processing at the SSC.  We
detect very weak latent images after changing the position of
HD\,189733, but their maximum amplitude ($\sim 0.002$ of the real
image) is not sufficient to account for the ramp.  It should eventually
be possible to diagnose the nature of this ramp, and perhaps correct
it from first principles.  However, the secondary eclipse is of
immediate interest, so here we fit a polynominal to the baseline.

The limiting factor in the accuracy of our secondary eclipse
measurement is our ability to correct the ramp, and establish an
accurate photometric baseline spanning the eclipse.  The ramp has a
shape reminiscent of $y \propto ln(\delta t)$, where $\delta t$ is the
elapsed time from start of observations, but a first-order log
function does not fit it particularly well.  Moreover, the shape of
the ramp is slightly different for sources of different brightness
(background, star).

To correct the baseline, we first divide background-subtracted
HD\,189733 photometry by the comparison star.  We do not subtract
background from the comparison star, so that it will be a closer match
to the intensity level of HD\,189733 and also will have greater SNR.
It is still necessary to smooth the comparison star photometry, which
we do by fitting a fourth-order polynominal in $\ln(\delta t)$
(Figure~1).  Dividing HD\,189733 by this fit removes virtually all of
the higher-order curvature from the HD\,189733 time series, for phases
greater than $0.45$ (neglecting the strongly varying initial
measurements).  A residual ramp remains in HD\,189733 after the
division, but it is nearly linear.  We zero-weight the eclipse itself
($0.482\le \phi \le 0.518$), and we fit both a linear and a quadratic
baseline to the residual ramp in the HD\,189733 data.

Our baseline correction has the distinct advantage of not fitting a
higher-order polynominal directly to HD\,189733 data.  Because the
eclipse itself must be zero-weighted, the higher-order coefficients
would have to be fit `over the gap', a less robust procedure that we
avoid.  For investigators who wish to do their own baseline
corrections, we include an electronic table (Table~1) giving our
photometry both before and after baseline correction.  The original
data are freely available from the Spitzer Science Center.

\section{Results and Discussion}
\label{sec:disc}

\subsection{Stellar Flux}

Our background-subtracted photometry yields a stellar flux of
$127\pm8$ mJy for HD\,189733 at $16~\mu$m. We included an aperture
correction of 14\%, calculated using a $24~\mu$m MIPS PSF\footnote{see
{\tt http://ssc.spitzer.caltech.edu/mips/psf.html}} scaled to
$16~\mu$m.  Our flux error is from calibration scatter described in
the IRS Data Handbook\footnote{{\tt
http://ssc.spitzer.caltech.edu/irs/dh/}}.  Interpolating in a grid of
Kurucz model atmospheres, the flux expected from a 5050/4.5/0.0 model
\citep{bouchy} at 19.3 pc is 104~mJy.  The 2MASS K magnitude (=5.54)
used with the STAR-PET calculator on the SSC website, for spectral
type K2, predicts a stellar flux of 120~mJy. Given the difference
between the 16~$\mu$m fluxes expected from the 2MASS magnitude, and
from a Kurucz model for the \citet{bouchy} temperature, there is no
convincing evidence for a circumstellar dust contribution to the
$16~\mu$m flux.  We therefore interpret the contrast in the secondary
eclipse solely in terms of the planet-to-star flux ratio.

\subsection{Amplitude of the Secondary Eclipse}

Figure~2 shows the baseline-corrected secondary eclipse.  We generate
a theoretical eclipse curve numerically \citep{richardson}, from the
\citet{bouchy} parameters for the system.  We fit this to the
individual measurements (upper panel of Figure~2).  The fit has only
two free parameters: phase at center of eclipse, and the eclipse
depth.  We estimate errors by generating and fitting to $10^4$ fake
data sets having $\sigma=0.0025$, matching the scatter in the real
data.  The standard deviations of the eclipse depth and central phase
from these Monte-Carlo realizations are adopted as the errors for
these parameters.  The best-fit eclipse depth is $0.551\%\pm0.017\%$,
with central phase $0.5026\pm0.0003$.  Adopting a stellar flux of
$\sim 120$ mJy, the flux from the planet is $\sim 660~\mu$Jy.

The exact eclipse depth is dependent on the details of the baseline
correction.  If we adopt a linear fit over the gap (see above), the
derived eclipse depth is $0.521\%$.  The choice between the linear and
quadratic baselines is largely subjective; we prefer the quadratic
baseline.  The magnitude of the difference in eclipse depth for the
two choices is indicative of the accuracy of our result, estimated as
$\pm0.03\%$.  This being greater than the precision of the detection,
baseline correction is the limiting factor in our analysis.

The lower panel of Figure~2 averages the photometry into bins of 0.001
in phase.  The eclipse in the binned data is dramatic, and its
duration and shape agree well with the theoretical curve.  We are
aware that some ground-based transit photometry for this planet yields
a smaller radius (D. Charbonneau and G. Bakos, private communication).
However, the duration of secondary eclipse should be the same as the
transit duration, already relatively well determined by the discovery
observations.  We do not expect that a smaller radius for the planet
will affect the duration and shape of the secondary eclipse curve by
more than our errors.  The brightness temperature inferred for the
planet may be more sensitive to radius (see below).

The central phase of the eclipse, $0.5026\pm0.0003$, is seemingly
different from $0.5$, and at face value this would indicate a
non-circular orbit.  However, the difference between the discovery
orbital period, 2.219 days \citep{bouchy}, and the Hipparcos value we
used (2.218575 days, \citealp{hebrard, bouchy}), when propagated to
the time of our observations, give a phase difference ($0.0046$) that
is greater than the offset we observe.  Therefore we regard any
conclusions about the central phase of the eclipse as premature until
the ephemeris is more firmly established.  For reference, our
secondary eclipse is centered at $HJD=2453692.62416\pm0.00067$.
Unlike the eclipse depth, the center time does not depend
significantly on our baseline correction.

It is interesting to compare this planet with the Spitzer detection of
TrES-1 \citep{charb05}, since both planets orbit K dwarfs. In thermal
equilibrium, and assuming equal Bond albedos and heat
re-distribution efficiencies for both planets, their temperatures will
scale as:

\begin{equation}
T_p \sim T_s \Delta^{\frac{1}{2}},
\end{equation}

\noindent where $T_s$ is the effective temperature of the star, and
$\Delta$ is its angular diameter as seen from the planet.  For the
TrES-1 \citep{alonso, sozetti} and HD\,189733 parameters
\citep{bouchy}, Eq. (1) predicts virtually identical temperatures, as
the larger and hotter star has the more distant planet (TrES-1). The
$16~\mu$m brightness temperature of HD\,189733, from a Kurucz model,
is 4315K.  Adopting the \citet{bouchy} ratio of radii, our secondary
eclipse contrast translates to a $16~\mu$m brightness temperature for
the planet of $1117\pm42$K.  This is very similar to both the TrES-1
($1060\pm50$K, \citealp{charb05}) and HD\,209458b ($1130\pm150$K,
\citealp{deming}) detections. Our contrast measurement is in close
accord with a prediction for HD\,189733b by \citet{fortney06}, who
expect HD\,189733b to be intermediate between TrES-1 and
HD\,209458b. However, if the planet-to-star radius {\it ratio} for
HD\,189733 is revised, then our revised brightness temperature will
be:

\begin{equation}
T^{16}_p = 900/\ln(1+41.86r^2), 
\end{equation}

\noindent where $r$ is the ratio of planet radius to the stellar radius. Our
$16~\mu$m measurement is as close to a continuum flux as possible
using Spitzer photometry; comparison with the IRAC and MIPS
($24~\mu$m) bands (DDT program 261 by D. Charbonneau) should define
the degree of absorption by methane, water and CO \citep{seager05,
fortney06}.

\subsection{Beyond the Eclipse Amplitude}

The fact that Spitzer detects this secondary eclipse to high SNR
($32\sigma$ precision) prompts us to ask what other information can be
extracted from the eclipse curve, and what are the implications for
the detection of lower-mass planets, e.g., hot Neptunes
\citep{bonfils} that may transit.

The lower panel of Figure~2 suggests that our binned data during
ingress and egress contain information on the shape of the eclipse
curve.  The pressure scale height in the atmosphere of a solar-type
star \citep{vernazza} is much less than the radius of a planet.
Hence, the shape of the secondary eclipse curve encodes information on
the spatial distribution of IR intensity across the planet's disk
\citep{charb05}.  We computed the derivative of intensity with respect
to time (phase) for our binned data.  We approximate the derivatives
as simple finite differences. We fold the eclipse curve about
mid-eclipse, averaging ingress and egress, before computing
derivatives. Folding maximizes our sensitivity, but smears any
planetary thermal structure that is not symmetric.

Figure~3 shows the results for two bin sizes: 0.001 in phase (upper
panel), and 0.002 (lower panel).  At 0.001 resolution (=~3~minutes)
the scatter is much greater than the amplitude of the derivative
curve, but the average derivative during ingress/egress is higher than
at other phases.  As we increase the bin size, the derivatives during
ingress/egress must inevitably become significantly different from
zero. This follows because the integral of the derivative is the
eclipse curve itself, that is already detected to high significance.
At 0.002 resolution (6~minutes), the scatter is dramatically improved,
the derivatives peak near mid-ingress/egress, and a $\chi^2$ analysis
firmly rejects the null hypothesis that the derivatives are consistent
with zero ($10^{-5}$ probability). A similar analysis for the
comparison star finds only random noise.  Apart from the real
derivative increase during ingress/egress, only the lower left bin on
Figure~3 stands out in the $\chi^2$ analysis, with a $5\%$ probability
of being due to noise.  This marginal value is due to the extra noise
noticeable just before egress, rather than to a real derivative
signal. The extra noise can also be seen in the background flucuations
at this time (see Figure~1).

Figure~3 compares the observed derivatives to models that are
constrained to fit the observed eclipse depth. We use two simple,
ad-hoc models: a uniform brightness temperature on the disk (solid
line), and an extreme limb-darkened model with intensity falling
proportional to $\cos(\theta)$, becoming zero at the limb (dashed
line). The dominant effect in the modeled and observed derivatives is
the shape of the planet. At 6-minute resolution, the individual
derivatives peak at the middle of ingress/egress.  The data crudely
indicate the overall shape of the planet, but cannot discriminate
between a uniform disk (solid line) and the extreme limb-darkened
model (dashed line).  However, modest improvements in Spitzer duty
cycle and SNR may allow us to relax our symmetry assumption, and place
meaningful limits on dynamically-forced temperature asymmetries of
large spatial scale on the planet's disk \citep{cho, cooper}.

The $32\sigma$ precision we obtain for this secondary eclipse remains
a $> 3\sigma$ detection for a planet an order of magnitude smaller in
area, but in an identical orbit.  This limit corresponds to
$\sim$~1~Neptune radius. As we proceed down the main sequence,
planet-to-star contrast increases as the inverse square of the stellar
radius. For planets orbiting M~dwarfs \citep{bonfils, rivera, butler},
it follows from Eq.~(1) that, in the Rayleigh-Jeans limit, this
dominates the reduced heating of the planets.  If close-in Neptunes
are discovered transiting nearby M~dwarfs, their IR emission should
be detectable by Spitzer.
 
\acknowledgments This work is based on observations made with the
Spitzer Space Telescope, which is operated by the Jet Propulsion
Laboratory, California Institute of Technology under a contract with
NASA. Support for this work was provided by NASA.  We thank Gordon
Squires for his rapid handling of our proposal, and the entire Spitzer
staff for their efficient scheduling, and rapid data processing.  We
are grateful to Dave Charbonneau and Gaspar Bakos for discussions
concerning this planet, and we thank Daniel Devost, Kevin Uchida,
James Houck, and Gregory Sloan for their thoughts regarding IRS
systematics.  We thank an anonymous referee for comments that
improved this paper.

\clearpage

\begin{figure}
\begin{center}
\includegraphics[scale=0.9]{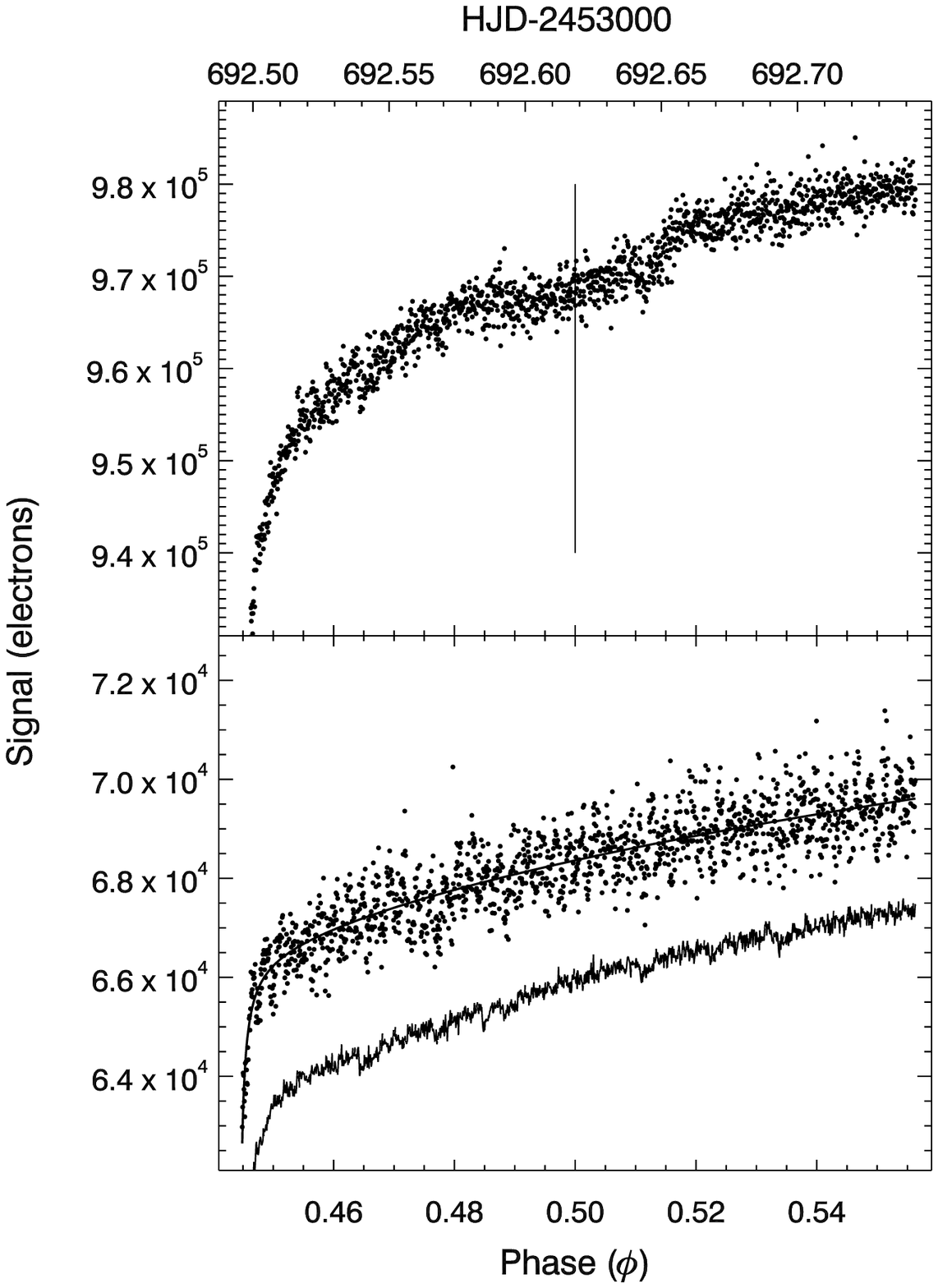}
\caption{ \emph{Upper Panel:} Raw aperture photometry, before
background subtraction and baseline correction, for HD\,189733 versus
planetary phase.  Note that the secondary eclipse is already visible
near phase 0.5 - marked by the vertical line.  \emph{Lower Panel:}
Aperture photometry of the comparison star (2MASS20004297+2242342,
points), with a polynominal fit (solid line through points, see text).
The line below the comparison star shows the background level, which
has been increased by an arbitrary factor to place it on the same
scale.  The background in the HD\,189733 aperture is about 30\% of the
total signal, and for the comparison star about 60\%.
\label{fig:raw} }
\end{center}
\end{figure}

\clearpage

\begin{figure}
\begin{center}
\includegraphics[scale=0.90]{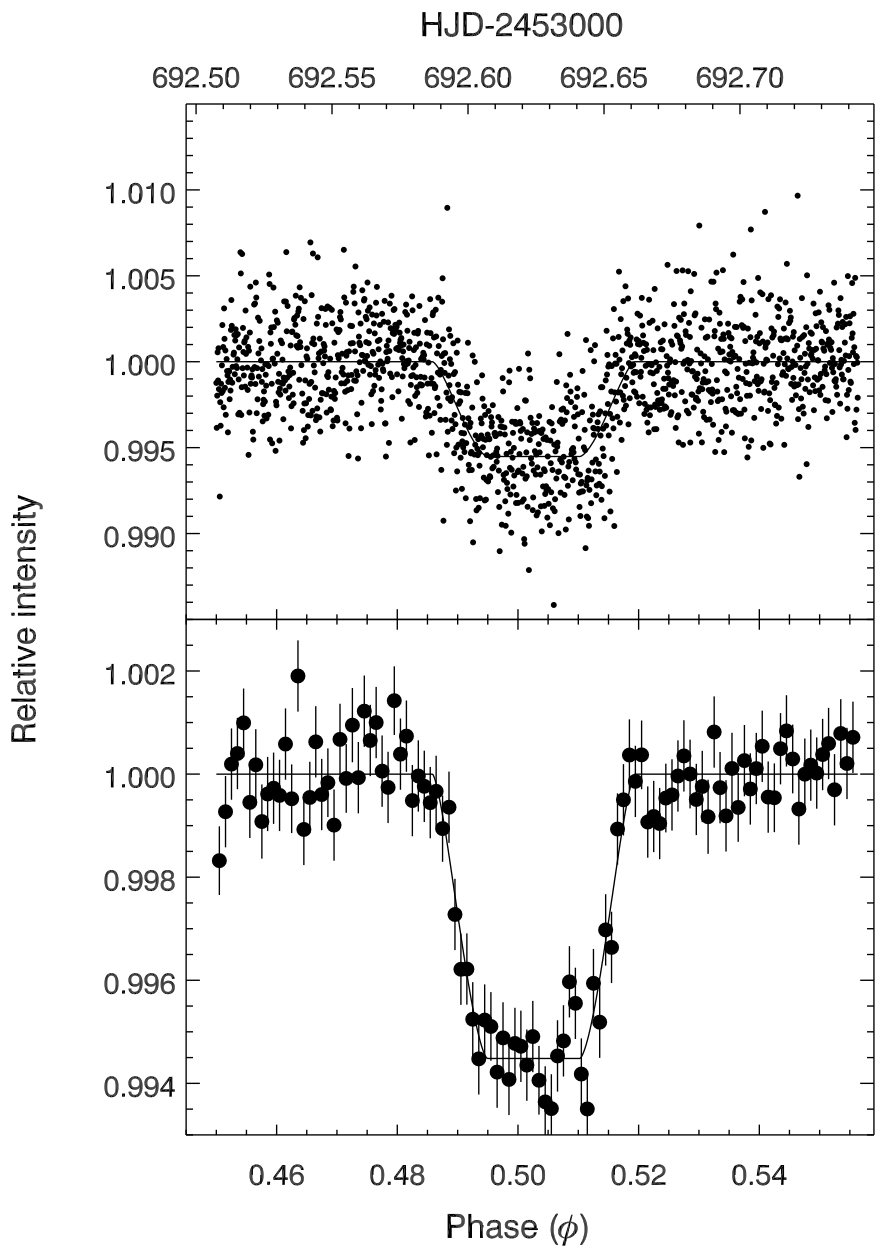}
\caption{ \emph{Upper Panel:} Baseline-removed aperture photometry of
the HD\,189733 secondary eclipse. Points are individual 6-second
measurements, with error bars supressed for clarity, but showing the
eclipse curve having the best-fit amplitude ($0.551\pm0.03\%$) and
central phase.  \emph{Lower Panel:} Data from the upper panel averaged
in bins of width 0.001 in phase ($\sim~3$~minutes), with error bars and the
best-fit eclipse curve.
\label{fig:phot} }
\end{center}
\end{figure}

\clearpage

\begin{figure}
\begin{center}
\includegraphics[scale=0.80]{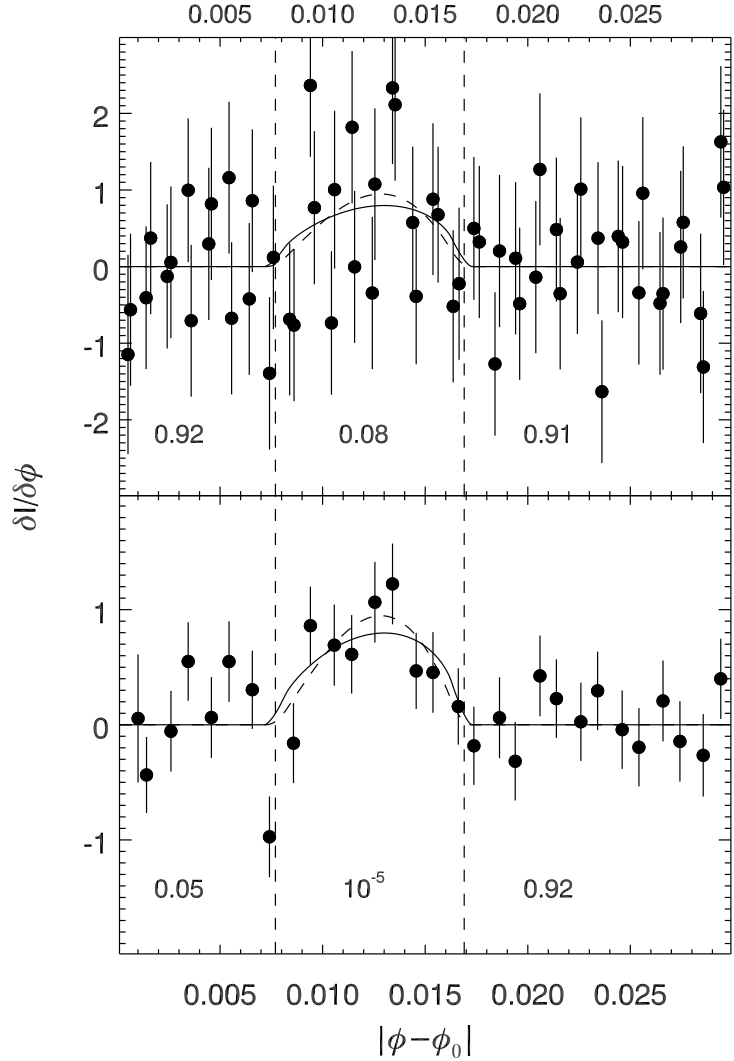}
\caption{ \emph{Upper Panel:} Derivative of relative intensity during
eclipse with respect to phase ($\phi$), versus the absolute phase
difference from the center of eclipse ($\phi_0$).  The derivative
values from the data were obtained using a phase resolution of
0.001.  The dashed vertical lines indicate the start
and end of ingress and egress.  \emph{Lower Panel:} Derivative of
relative intensity versus phase, as in upper panel, but for a phase
resolution of 0.002. The curved line shows the relation expected for a
circular planet of uniform brightness temperature, and the dashed line
has a temperature distribution sharply peaked at the planet's disk
center.  The numbers in each region of the figure are the
probabilities, from a $\chi^2$ analysis, that the observed derivatives
are consistent with zero.
\label{fig:deriv} }
\end{center}
\end{figure}

\clearpage

\begin{deluxetable}{lllll}
\tablecaption{Photometry before and after baseline correction. The
complete version of this Table is in the electronic edition 
of the Journal.  The printed edition contains only a sample.} 
\tablewidth{0pt} \tablehead{ \colhead{ } &
\colhead{HJD-2453000} & \colhead{Phase} & \colhead{Signal (electrons)}
& \colhead{Relative intensity}\tablenotemark{a}} \startdata 1 &
692.49608 & 0.44488 & 609247 & 0.0 \\ 2 & 692.49624 & 0.44495 & 610846
& 0.0 \\ 3 & 692.49639 & 0.44502 & 611768 & 0.0 \\ \enddata
\tablenotetext{a}{Baseline-corrected, but given only for phase greater
than 0.45}
\end{deluxetable}

\end{document}